%% file: K2-19_3to2.tex
\documentclass[a4paper,fleqn,usenatbib]{mnras}

\usepackage{newtxtext,newtxmath}
\usepackage[T1]{fontenc}
\usepackage{ae,aecompl}

\usepackage{graphicx}	%
\usepackage{amsmath}	%
\usepackage{booktabs}
\usepackage{xspace}

\newcommand{\theres}{\theta_{\mathrm{res}}}
\newcommand{\Gr}{\mathcal{G}}
\renewcommand{\i}{\iota}
\renewcommand{\H}{\mathcal{H}}
\newcommand{\tr}{\tilde{\mathbf{r}}} 
\renewcommand{\r}{\mathbf{r}} 
\newcommand{\R}{\mathcal{R}}
\newcommand{\lc}[3]{b_{#1}^{#2}(#3)}
\renewcommand{\epsilon}{\varepsilon}

\graphicspath{{./Figures/}}

\newcommand{\Mstar}{\ensuremath{M_{\star}}\xspace}
\newcommand{\Rstar}{\ensuremath{R_{\star}}\xspace}

\newcommand{\Me}{\ensuremath{M_{\oplus}}\xspace} 
\renewcommand{\Re}{\ensuremath{R_{\oplus}}\xspace}

\newcommand{\Rsun}{\ensuremath{R_{\odot}}\xspace }
\newcommand{\Msun}{\ensuremath{M_{\odot}}\xspace}

\def\deg{\ensuremath{^{\circ}}}

\usepackage{xstring} %
\usepackage{ifthen} %
\usepackage{xifthen} %
\newcommand{\argperi}[1]{\ensuremath{\ifthenelse{\isempty{#1}}{\omega_P}{\omega_{P,#1}}}\xspace}
\newcommand{\inc}[1]{\ensuremath{\ifthenelse{\isempty{#1}}{i}{i_{#1}}}\xspace}
\newcommand{\ecc}[1]{\ensuremath{\ifthenelse{\isempty{#1}}{e}{e_{#1}}}\xspace}
\newcommand{\per}[1]{\ensuremath{\ifthenelse{\isempty{#1}}{P}{P_{#1}}}\xspace}
\newcommand{\node}[1]{\ensuremath{\ifthenelse{\isempty{#1}}{\Omega}{\Omega_{#1}}}\xspace}
\newcommand{\meananom}[1]{\ensuremath{\ifthenelse{\isempty{#1}}{M}{M_{#1}}}\xspace}
\newcommand{\ecosw}[1]{\ensuremath{\ifthenelse{\isempty{#1}}{e \cos \omega_P}{e \cos \omega_{#1}}}\xspace}
\newcommand{\esinw}[1]{\ensuremath{\ifthenelse{\isempty{#1}}{e \sin \omega_P}{e \sin \omega_{#1}}}\xspace}
\newcommand{\T}[1]{\ensuremath{\ifthenelse{\isempty{#1}}{T}{T_{#1}}}\xspace}
\newcommand{\tc}[1]{\ensuremath{\ifthenelse{\isempty{#1}}{T_{c}}{T_{c,#1}}}\xspace}
\newcommand{\secosw}[1]{\ensuremath{\sqrt{e_{#1}} \cos \omega_{#1}}\xspace}
\newcommand{\sesinw}[1]{\ensuremath{\sqrt{e_{#1}} \sin \omega_{#1}}\xspace}
\newcommand{\rrat}[1]{\ensuremath{R_{p,#1}/\Rstar}\xspace}

\input{val_sys.tex}

\title[The resonant dynamics of the K2-19 system]{Resonance in the K2-19 system is at odds with its high reported eccentricities}

\author[A. C. Petit et al.]{Antoine C. Petit$^{1,2,}$\thanks{antoine.petit@astro.lu.se}, Erik A. Petigura$^{3}$, Melvyn B. Davies$^{1}$ and Anders Johansen$^{1}$\\
	$^{1}$ Lund Observatory, Department of Astronomy and Theoretical Physics, Lund University, Box 43, 22100 Lund, Sweden\\
	$^{2}$ IMCCE, CNRS-UMR8028, Observatoire de Paris, PSL University, Sorbonne Université, 77 Avenue Denfert-Rochereau, 75014 Paris, France\\
	$^{3}$ Department of Physics \& Astronomy, University of California Los Angeles, Los Angeles, CA 90095, USA
}

\date{Accepted XXX. Received YYY; in original form ZZZ}

\pubyear{2020}

\begin{document}
	\label{firstpage}
	\pagerange{\pageref{firstpage}--\pageref{lastpage}}
	\maketitle

\begin{abstract}
	K2-19 hosts a planetary system composed of two outer planets, b and c, with size of $7.0\pm 0.2~\Re$ and $4.1\pm0.2~\Re$ , and an inner  planet, d, with a radius of $1.11\pm 0.05 \Re$.
	A recent analysis of  Transit-Timing Variations (TTVs) suggested b and c are close to but not in 3:2 mean motion resonance (MMR) because the classical resonant angles circulate.
	Such an architecture challenges our understanding of planet formation. Indeed, planet migration through the protoplanetary disc should lead to a capture into the MMR. Here, we show that the planets are in fact, locked into the 3:2 resonance despite circulation of the conventional resonant angles and aligned periapses.
    However, we show that such an orbital configuration cannot be maintained for more than a few  hundred million years due to the tidal dissipation experienced by planet d.
    The tidal dissipation remains efficient because of a secular forcing of the innermost planet eccentricity by planets b and c.
    While the observations strongly rule out an orbital solution where the three planets are on close to circular orbits, it remains possible that a fourth planet is affecting the TTVs such that the four planet system is consistent with the tidal constraints.
\end{abstract}
	
\section{Introduction}

The numerous planet discoveries over the past decades have revealed the large diversity in sizes and orbital architecture of exoplanetary systems \citep{Winn2015}.
The discovery of Hot Jupiters \citep{Mayor1995} and of super-Earths have profoundly changed how we see planets and their formation outside of the Solar System narrative.
Since then, planet formation theories have been adapted to take this diversity into account.
The models now prefer a fast formation within the protoplanetary disc lifetime involving a migration process \citep[\emph{e.g.}][and references therein]{Bitsch2019,Izidoro2019,Lambrechts2019}.
In particular, the migration process can lead to the capture of planet in mean motion resonant (MMR) chains \citep{Cresswell2008}.
Most of the resonant chains are expected to break once the protoplanetary disc is dissipated as most of the systems are observed outside of MMR \citep{Izidoro2017,Izidoro2019}.
Nevertheless, the period ratio distribution still shows signpost of this past history \citep{Fabrycky2014}.
The study of the few remaining resonant chains is thus of particular interest to unravel the very early life of planetary systems.

The capture into MMR is a complex process that depends on the migration parameters, the planet masses and the particular resonance where capture happen \citep[\emph{e.g.}][]{Mustill2011,Batygin2015}.
Nevertheless, we expect planets in resonant chains to have close to circular orbits due to the eccentricity damping during the migration \citep{Cresswell2008}.
Such configurations have been observed for various systems observed through both radial velocities (RV) and with the analysis of Transit-Timing Variations \citep[TTV, see \emph{e.g.},][]{Agol2005,Holman2005}. 
Systems where observations challenge this theoretical picture are of particular interest as they are the only way to probe the validity and/or the generality of theories.

K2-19 hosts three known transiting planets. The first two were reported  \citep{Armstrong2015} based on the photometry collected by the \emph{Kepler Space Telescope} during the \emph{K2} operations \citep{Howell2014}. K2-19 b and c appeared to be close to the 3:2 MMR but it was not possible to conclude whether or not the pair was indeed in resonance. 
Planets b and c  sizes lies between  Uranus and Saturn with respective radii of $7.0\pm0.2~\Re$ and $4.1\pm0.2~\Re$ and they orbit in respectively 7.9 d and 11.9 d.
A third inner planet was detected by \cite{Sinukoff2016}.
Planet d orbits in 2.5 d and has a size similar to Earth with a radius of ${1.11\pm0.05 \Re}$.
The recent observations by \cite{Petigura2020} of TTV for both planets b and c, as well as Radial Velocities (RV) measurements showing the reflex motion due to planet b, have given a precise set of orbital elements for the system.
The 5\% fractionnal uncertainties, are among the smallest  for sub-Jovian exoplanets.
Using a photodynamical model, they obtained a determination of the planet masses and orbital element down to a few percent.
One puzzling aspect of this system is the moderate eccentricities of 0.2 with well-aligned apsides ($\Delta \varpi = 2\pm2~\deg$) of planets b~and~c.
\cite{Petigura2020} conclude from these observations that the system is very close to the 3:2 commensurability while not being resonant based on the analysis of the classical resonant angles. 
K2-19 system's architecture is thus puzzling from a dynamical point of view as no clear mechanism is identified to explain its stability. 
Indeed, \cite{Petigura2020} report that the system is stable in numerical simulations but is strongly AMD-unstable \citep{Laskar2017} and not protected by the resonance.
Moreover, such a configuration is in tension with our current understanding of planet formation.
Indeed, convergent migration within the protoplanetary disc leads to eccentricity damping and capture into MMR \citep{Cresswell2008}. The planets trapped into such state have low eccentricity (comparable to the protoplanetary disc aspect ratio $\simeq 0.05$) and have their periapses anti-aligned \citep{Batygin2013}.

The dynamics of the system and its origin thus necessitate an in-depth study. 
From a dynamical point of view, the proximity to the resonance and the eccentric and aligned orbits  require one to go beyond the simple study of the resonant angles as they may not be representative of the nature of the dynamics.
Given that all the planets are within 0.1 au, tidal effects must also be considered.

In this paper, we revisit the dynamical study of the system orbiting K2-19.
We show in section \ref{sec:resandsec} that the two outer planets are indeed trapped into the 3:2 mean motion resonance despite being apsidally aligned and the resonant angles circulation.
The first-order resonant model explains all the dynamics properties discussed by \cite{Petigura2020}.
We also show that the inner planet is secularly coupled to the b-c pair.
We then study in section \ref{sec:tides} the effect of tidal dissipation onto the inner planet d.
We show that due to the eccentricity forcing from the outer planets, planet d's orbit tends to decay while the outer planets circularize.
The time scale for the system evolution is shorter than its lifetime rendering the configuration unlikely.
In these two sections, we take the results from \cite{Petigura2020} as certain and simply draw conclusions based on the dynamical analysis.
We finally discuss in section \ref{sec:bias} the tension between the observations and our theoretical understanding of the system history. In particular we highlight the constraints on the three planet best fit and discuss whether the TTVs might be affected by an unmodelled effect, including the presence of a planet not yet detected.

\section{Resonant dynamics of K2-19's system}
\label{sec:resandsec}

In this section, we re-analyse the dynamics of the best-fitting three planet solution given by \cite{Petigura2020}.
We show that the outer planets are indeed inside the MMR and that they are coupled secularly to the inner planet. 
We partially reproduce in  Table  \ref{tab:orb3pl}, the orbital elements and planet characteristics from the  best photodynamical fit from \cite{Petigura2020}.

\begin{table}
	\begin{center}
	\caption{K2-19 system parameters from \citep{Petigura2020}, reproduced by permission of the AAS.\label{tab:orb3pl}}
\begin{tabular}{lr}
	\hline
	\hline
	Parameter              & Value    \\
	\hline
	\Mstar (\Msun)            & \sys{pd-mstar_fmt}        \\
	\Rstar (\Rsun)            & \sys{pd-rstar_fmt}     \\
	$P_b$ (d)              & \sys{pd-per2_fmt} \\
	$\tc{b}$ (BJD$-$2454833)  & \sys{pd-tc2_fmt}   \\
	$\secosw{b}$              & \sys{pd-secosw2_fmt}  \\
	$\sesinw{b}$              & \sys{pd-sesinw2_fmt}   \\
	$\inc{b}$ (deg)           & \sys{pd-inc2_fmt}  \\
	$\Omega_b$ (deg)          & 0 (fixed)     \\
	$\rrat{b}$                & \sys{pd-ror2_fmt} \\
	$M_{p,b}$ (\Me)           & \sys{pd-masseprec2_fmt}   \\
	$P_c$ (d)              & \sys{pd-per3_fmt}      \\
	$\tc{c}$ (BJD$-$2454833)  & \sys{pd-tc3_fmt}        \\
	$\secosw{c}$              & \sys{pd-secosw3_fmt}      \\
	$\sesinw{c}$              & \sys{pd-sesinw3_fmt}      \\
	$\inc{c}$ (deg)           & \sys{pd-inc3_fmt}         \\
	$\Omega_c$ (deg)          & \sys{pd-Omega3_fmt}    \\
	$\rrat{c}$                & \sys{pd-ror3_fmt}   \\
	$M_{p,c}$ (\Me)           & \sys{pd-masse3_fmt}   \\
	$P_d$ (d)              & \sys{pd-per1_fmt}     \\
	$\tc{d}$ (BJD$-$2454833)  & \sys{pd-tc1_fmt}       \\
	$\secosw{d}$              & 0 (fixed)                 \\
	$\sesinw{d}$              & 0 (fixed)            \\
	$\inc{d}$ (deg)           & \sys{pd-inc1_fmt}        \\
	$\Omega_d$ (deg)          & 0 (fixed)                \\
	$\rrat{d}$                & \sys{pd-ror1_fmt}         \\
	$M_{p,d}$ (\Me)           & <10                      \\
	\multicolumn{2}{l}{{\bf Derived Parameters}} \\
	$R_{p,b}$ (\Re)           & \sys{pd-prad2_fmt}        \\
	$R_{p,c}$ (\Re)           & \sys{pd-prad3_fmt}       \\
	$R_{p,d}$ (\Re)           & \sys{pd-prad1_fmt}       \\
	$\ecc{b}$                 & \sys{pd-e2_fmt}          \\
	$\ecc{c}$                 & \sys{pd-e3_fmt}          \\
	$\Delta \omega$ (deg)     & \sys{pd-omegadiffdeg_fmt} \\
	\hline
	\\[-6ex]
\end{tabular}
\end{center}
\end{table}

\subsection{The 3:2 mean motion resonance}	
\label{sec:32res}

The analysis of the K2 photometric data makes it clear that K2-19 b and c are close to the 3:2 MMR.
Being close to the 3:2 MMR means that the planet mean motions $n_k=2\pi/P_k$ ($P_k$ being the planets' orbital periods) satisfy the arithmetic relation $2 n_b -3n_c\simeq 0$ .
As a result, the motion of the planets are coupled and one cannot average over the fast motions to study the long-term orbital evolution.
Instead, the classical approach to analyse resonant motions consists of averaging the planet interactions over the non-resonant angles to reduce the problem to a one degree of freedom problem that is integrable.
For first-order MMR such as  the 3:2 resonance, d'Alembert relations \citep[see][]{Morbidelli2002} impose that at first-order, the resonant terms in the development of the perturbation depend on the angles
\begin{equation}
\varphi_k = 2\lambda_b-3\lambda_c+\varpi_k
\label{eq:resanglesclassic}
\end{equation}
where $\lambda_k$ and  $\varpi_k$  are respectively the mean longitude  and the longitude of the periapsis of planet $k$. 
There are two different combination of angles related to the 3:2 resonance.
In principle, the interaction between the two terms should make the system not integrable.
In reality, the system can be reduced to a one degree of freedom resonant system thanks to a constant of motion that appears after a canonical transformation \citep{Sessin1984,Henrard1986}.

The integrable approximation for first-order MMR has been called the second fundamental model of resonance\footnote{The first fundamental model is the classical pendulum.} \citep{Henrard1983}.
The analytical derivation of the integrable model for two massive planets has been carried out by several authors \citep{Henrard1986,Batygin2013,Deck2013,Delisle2014a,Petit2017,Hadden2019}.
It is obtained by an expansion to first-order in eccentricity, averaging over the fast angle and a rotation of the two classical resonant coordinates 
\begin{equation}
x_k = \sqrt{C_k}e^{\i (\varpi_k-\theres )} \simeq \sqrt{\frac{\Lambda_k}{2}}e_ke^{\i  (\varpi_k-\theres )} 
\end{equation}
where ${C_k = \Lambda_k(1-\sqrt{1-e_k^2})}$ is the angular momentum deficit \citep[AMD,][]{Laskar1997} of the planet $k$, ${\Lambda_k = m_k\sqrt{\Gr m_S a_k}}$,  $\Gr$ is the gravitational constant and $\theres = 3\lambda_c-2\lambda_b$.
Following \citep{Petit2017}, we also define 
\begin{equation}
\tilde{e}_k = \sqrt{\frac{2C_k}{\Lambda_k}}=\sqrt{2\left(1-\sqrt{1-e_k^2}\right)} \simeq e_k.
\end{equation}
The rotation, first described in \cite{Sessin1984}, transforms the coordinates $x_b,x_c$ into two complex coordinates $y_1$ and $y_2$ (we follow here the notations from \citealp{Petit2017}).
The norm of $y_2$ is a constant of motion and the dynamics of $y_1$ are described by the second fundamental model of resonance \citep[see][for a complete description of the dynamics]{Ferraz-Mello2007}.
It is also worth pointing out that the total AMD of the system is given by $C=I_1+I_2$ where $I_k = y_k\bar{y}_k$.
For the 3:2 MMR, the expressions of $y_1$ and $y_2$ are 
\begin{align}
y_1 &= \sqrt{\frac{\Lambda_b}{2}} \frac{\tilde{e}_b e^{\i  \varpi_b }- 1.22 \tilde{e}_ce^{\i  \varpi_c }}{\sqrt{1+1.31\gamma} }e^{-\i\theres},\label{eq:y1}\\
y_2 &= \sqrt{\frac{\Lambda_c}{2}}\frac{1.07 \gamma \tilde{e}_b e^{\i  \varpi_b }+  \tilde{e}_c e^{\i  \varpi_c }}{\sqrt{1+1.31\gamma}}e^{-\i\theres},
\label{eq:y2}
\end{align}
where $\gamma=m_b/m_c\simeq3.0$ for K2-19 and the numerical coefficients come from the expansion of the resonant terms of the 3:2 resonance. More precisely, the coefficients are linear combinations of different Laplace coefficients evaluated at the exact Keplerian resonance. We give the analytical expression in appendix \ref{app:ham} and refer to \cite{Batygin2013,Petit2017} or \cite{Hadden2019} for recent complete derivations of the Hamiltonian.

\begin{figure}
	\includegraphics[width=\linewidth]{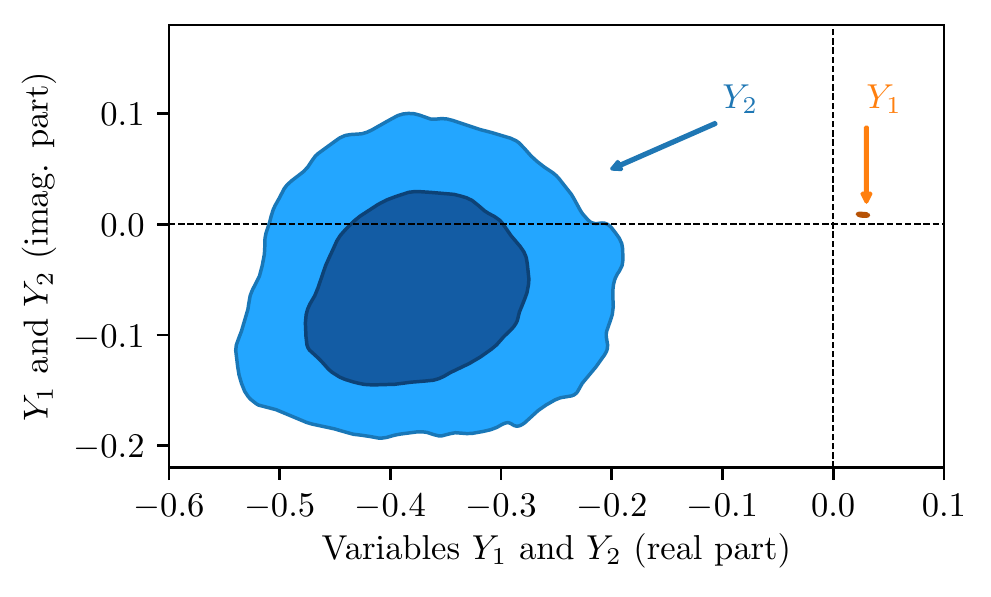}
	\caption{Posterior distribution of the real and imaginary part of $Y_1$ and $Y_2$ calculated from \citep{Petigura2020}. Colours correspond to the $1\sigma$ and $2\sigma$ confidence levels. The posterior distribution of $Y_1$ is plotted to show the differences in uncertainties.}
	\label{fig:Y12_hist}
\end{figure}

Without loss of generality, one may rescale $y_1$ and $y_2$
\begin{align}
Y_1 &= \sqrt{\frac{2}{\Lambda_b}}y_1 = \frac{\tilde{e}_b e^{\i  \varpi_b }- 1.22 \tilde{e}_ce^{\i  \varpi_c }}{\sqrt{1+1.31\gamma} }e^{-\i\theres},\label{eq:Y1}\\
Y_2 &= \sqrt{\frac{2}{\Lambda_c}}y_2 = \frac{1.07 \gamma \tilde{e}_b e^{\i  \varpi_b }+  \tilde{e}_c e^{\i  \varpi_c }}{\sqrt{1+1.31\gamma}}e^{-\i\theres}.
\label{eq:Y2}
\end{align}
They are linear combinations of the eccentricities where $Y_1$ is roughly the eccentricity vectors difference and $Y_2$ the  mass weighted sum.
The posterior distributions for $Y_1$ and $Y_2$ calculated from \cite{Petigura2020} are shown in Figure \ref{fig:Y12_hist}. As we can see, $Y_1$ is much more constrained than $Y_2$. 

\begin{figure}
	\includegraphics[width=8.5cm]{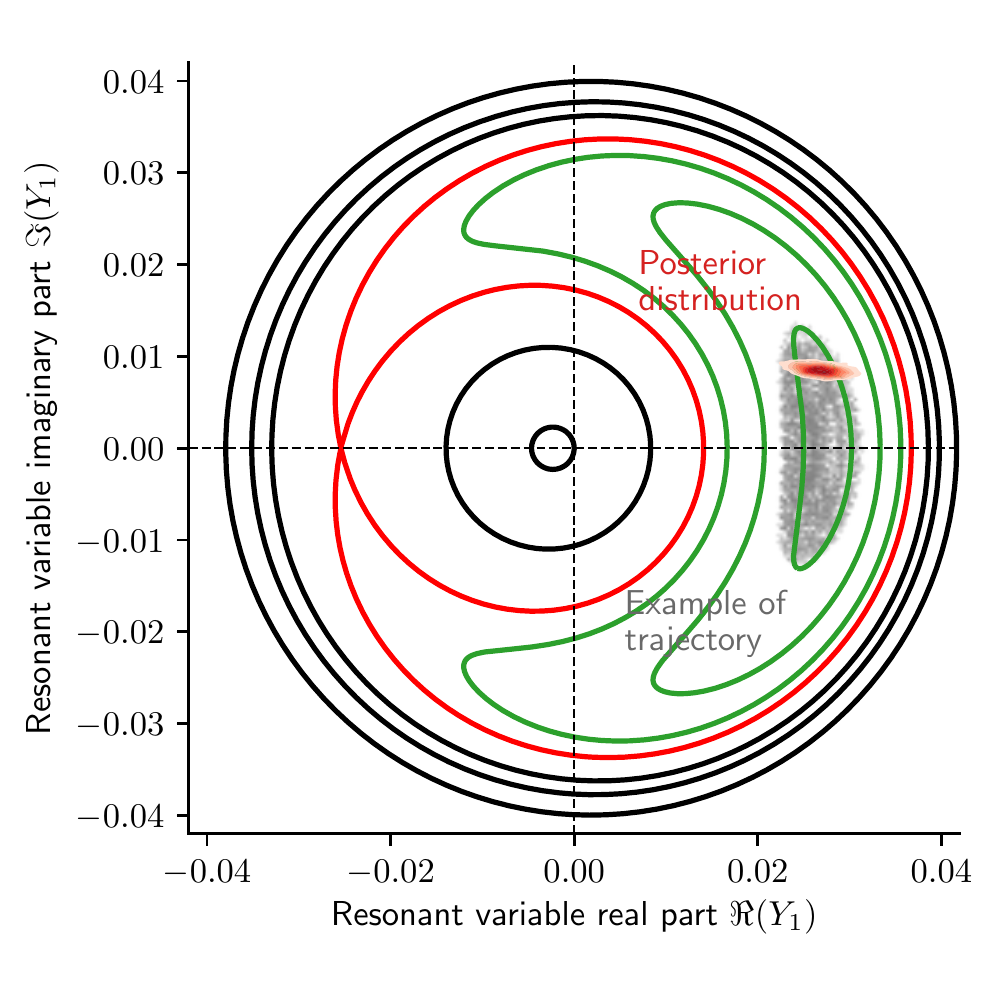}
	\caption{Phase portrait of the integrable approximate Hamiltonian in the $Y_1$ (eq. \ref{eq:Y1}) space. The black lines correspond to circulating orbits, the green lines to resonant orbits and the red line is the separatrix. The red set represents the posterior probability distribution of $Y_1$  at time $t=2020 \ \text{BJD}-2454833$.
	The grey points corresponds to the result of one $N$-body integration of the two planet system for 5000 planet b orbits.}
	\label{fig:phaseportrait}
\end{figure}

In the case of a system with two massive planets, the real resonant angle corresponds to the argument of $Y_1$.
Hence, to determine if the system is indeed in resonance one should in principle verify if the variable $Y_1$ evolves within the resonant island shown in figure \ref{fig:phaseportrait}.
In reality, for  most of the resonant chains observed in exoplanet systems, the resonant angles $\varphi_k$ (eq. \ref{eq:resanglesclassic}) are good proxies for the actual resonant angle and such an analysis is not needed (see below).

On the other hand, the actual value of $Y_2$ is less critical to determine whether or not  the system is resonant because it does not affect directly the shape of the resonance (fig \ref{fig:phaseportrait}). 
It can also be shown (see appendix \ref{app:ham}) that within the limit of the first-order model, $Y_2$ precesses at the same frequency as $\theres$. Hence $Y_2e^{\i\theres}$ is almost a constant that we will note
 \begin{equation}
 \tilde{Y}_2=Y_2e^{\i\theres} =\frac{1.07 \gamma \tilde{e}_b e^{\i  \varpi_b }+  \tilde{e}_ce^{\i  \varpi_c }}{\sqrt{1+1.31\gamma}}.
  \end{equation}
In reality, secular terms are neglected in this approximation. Nevertheless, the evolution of $\tilde{Y}_2$ happens on a much longer time scale.

The phase space dynamics of $Y_1$ are shown in Figure \ref{fig:phaseportrait}. We compute the phase portrait using the expression of the Hamiltonian given in the appendix (eq. \ref{eq:ham}). 
We also plot the distribution of $Y_1$ using the posterior distribution of \cite{Petigura2020}.
It should be noted that while the uncertainties on the eccentricities are of the order of 0.03 in this paper, $Y_1$ has a much more restricted spread, reinforcing the argument that the system is in a resonant state i.e., the data require that $Y_1$ is well-within the resonant island shown in Figure \ref{fig:phaseportrait}.
Finally we plot the result of a numerical integration of the two planet system.
We integrated multiple draws from the posterior distribution from \cite{Petigura2020} with \texttt{REBOUND} \citep{Rein2012a} and the high order integrator \texttt{SABA(10,6,4)} \citep{Blanes2013,Rein2019} and for 5000 orbits of planet b (roughly 108 years). We show a sample trajectory in Figure \ref{fig:phaseportrait}. All draws behaved in qualitatively the same way, i.e., showing the libration of the resonant variable $Y_1$.

It is clear that the dynamics are resonant, moreover, we can see that all the trajectories lie very close to the centre of the resonance. 
It suggests that the mechanism that led to the capture must have been gentle and dissipative \citep{Batygin2015}.
The most favoured mechanism for the formation is through migration within a protoplanetary disc \citep[\emph{e.g.}][]{Cresswell2008}. However, such a scenario is incompatible with the large eccentricities as well as the apsidal alignment of planet b and c as discussed in \cite{Petigura2020}.

\subsection{Short-term eccentricity evolution}
\label{sec:shorttermecc}

\cite{Petigura2020} also reported that $e_b$ and $e_c$ oscillate over a period of about 6 yr. We show that this oscillation is well explained by the resonant dynamics.

In most of the systems that have been observed in resonant configurations, $Y_2$ is usually negligible with respect to $Y_1$. 
When $|Y_2| \ll |Y_1|$, the libration of the angles $\varphi_k$ (eq. \ref{eq:resanglesclassic}) is a good proxy to test whether or not a system is in MMR.
However, in cases where $Y_2$ cannot be neglected with respect to $Y_1$, the transformation from this set of resonant variables to the classical orbital elements is not straightforward.
This means that the angles $\varphi_k$ can circulate while the system is actually very well described by the resonant first-order integrable model.
Indeed, the inverse transformation from $Y_k$ to the complex eccentricities can be written as
\begin{align}
\tilde{e}_b e^{\i (\varpi_b-\theres)} &= \frac{Y_1+1.22Y_2}{\sqrt{1+1.31\gamma}},\nonumber\\
\tilde{e}_c e^{\i (\varpi_c-\theres)} &=\frac{-1.07\gamma Y_1 +Y_2}{\sqrt{1+1.31\gamma}}\label{eq:invtransf}.
\end{align}
Since $Y_1$ oscillates around the resonant centre, its argument is librating. In a very rough approximation, we can consider it as constant.
On the other hand $Y_2$ has a constant norm and is rotating, with a frequency comparable to the one of $\theres$ that can be approximated as 
\begin{equation}
\nu_\theta = -3\delta n_c \simeq -2.7\times 10^{-3}\text{ rad.day}^{-1},
\end{equation}
where $\delta = 2n_b/(3n_c)-1$, represents the distance to the exact Keplerian resonance.
In this approximation, $\tilde{e}_b e^{\i (\varpi_b-\theres)}$ describes a circle centred on $Y_1/\sqrt{1+1.31\gamma}$ and of radius $1.22Y_2/\sqrt{1+1.31\gamma}$ (a similar analysis can be done for planet c).
The angle $\varphi_b=\varpi_b-\theres$ librates if the complex plane origin lies outside of this circle.
From eq.~\eqref{eq:invtransf}, we see that if $|Y_1|\la1.22|Y_2|$ (resp. $1.07\gamma|Y_1|\la|Y_2|$), then $\varphi_b=\varpi_b-\theres$ (resp. $\varphi_c=\varpi_c-\theres$) circulates.
In the case of the best fit studied here, both of these conditions are fulfilled.
As a result, we cannot use the classical angles to probe the resonance.
To our knowledge, this is the first system observed where the resonance cannot be characterized thanks to the classical angles.

\begin{figure}
	\includegraphics[width=\linewidth]{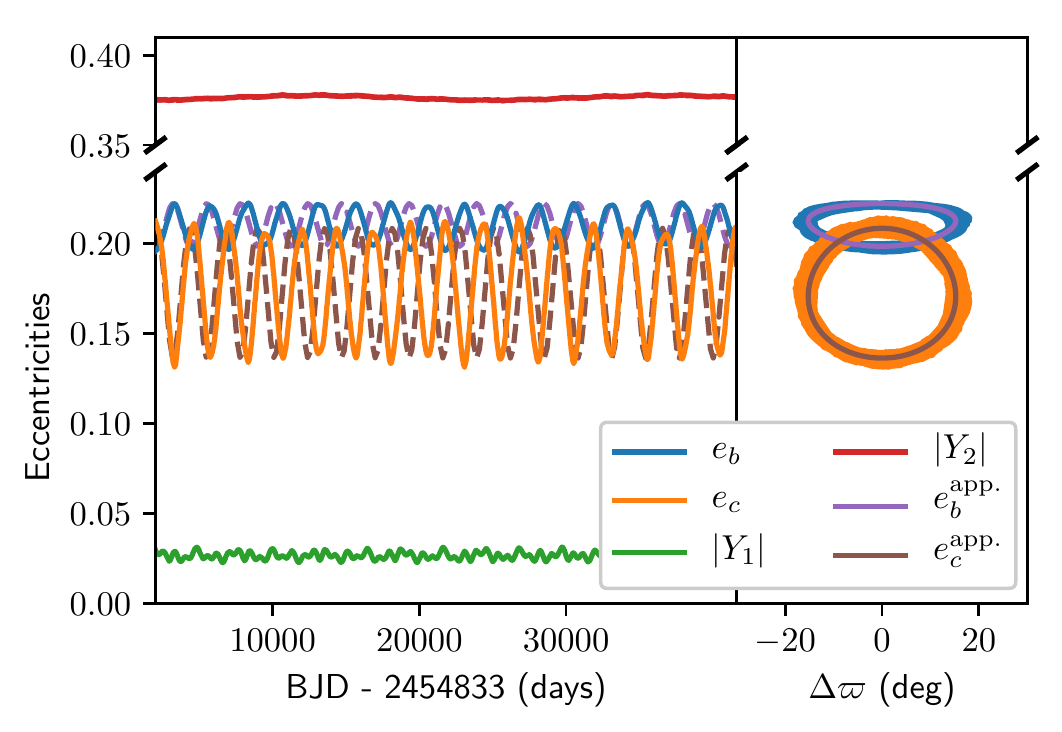}
	\caption{Dynamical evolution of the eccentricities and resonant variables of the numerical solution described in section \ref{sec:32res}. The quantities $Y_k$ are defined by eq. \eqref{eq:Y1} and \eqref{eq:Y2} and $e_k^{\text{app.}}$ by eq. \eqref{eq:eccevo}. \label{fig:eccevol2pl}}
\end{figure} 

The 3:2 resonance can also explain the eccentricity dynamics and in particular the apsidal aligned configuration.
Let us denote with  $Y_1^0$ and $Y_2^0$ the initial conditions for $Y_1$ and $Y_2$ at $t_0$, the evolution of the eccentricities can be approximated as
\begin{align}
e_be^{\i\varpi_b} &\simeq 0.46 Y_1^0 e^{-\i\nu_\theta(t-t_0)} + 0.56 Y_2^0 ,\nonumber\\
e_c e^{\i\varpi_c}&\simeq -1.40 Y_1^0 e^{-\i\nu_\theta(t-t_0)} + 0.46 Y_2^0 ,
\label{eq:eccevo}
\end{align}
where $\gamma$ was replaced by its average value and we used the approximation $\tilde{e}_k\simeq e_k$. 
From eq. \eqref{eq:eccevo}, we see that the oscillations of $e_c$ are about three times as large as the one of $e_b$. The period of oscillations should be of order 6.3 yr.

We plot in Figure \ref{fig:eccevol2pl} the evolution of the eccentricities of planet b and c, of $|Y_1|$ and $|Y_2|$ as well as the approximation from eq. \eqref{eq:eccevo}. While there is a small discrepancy on the frequency (the error is of the order of 10\%), the amplitude of the motion is well reproduced. Moreover, the agreement in the plane $e_k$-$\Delta\varpi$ is very good.

The main observational evidence for the resonance comes from the TTV. While the eccentricities evolve with a period of about 6 yr, the main frequency in the observed TTVs from \cite{Petigura2020} is about  2 years, which is the period of the oscillation of the resonant variable $Y_1$.

\subsection{Secular evolution}
 \label{sec:secular}
On short time-scales (\emph{i.e.} comparable to the TTV baseline), the simple model described above gives a good description of the system dynamics. 
On longer time-scales (more than a few kyr), the system is subject to orbital precession due to secular interactions.
We integrate the same initial condition as in the previous section, but this time we add planet d to the system.
The initial condition was drawn from the posterior distribution and the mass of planet d in this particular realization\footnote{The value is close to the average of the posterior distribution but results were qualitatively similar for other realizations (not shown here).} is 5.9 M$_\oplus$.
The simulation is run for 10,000 years, general relativity and stellar oblateness slightly change the precession rate but are not included in the example shown.

\begin{figure}
	\includegraphics[width=\linewidth]{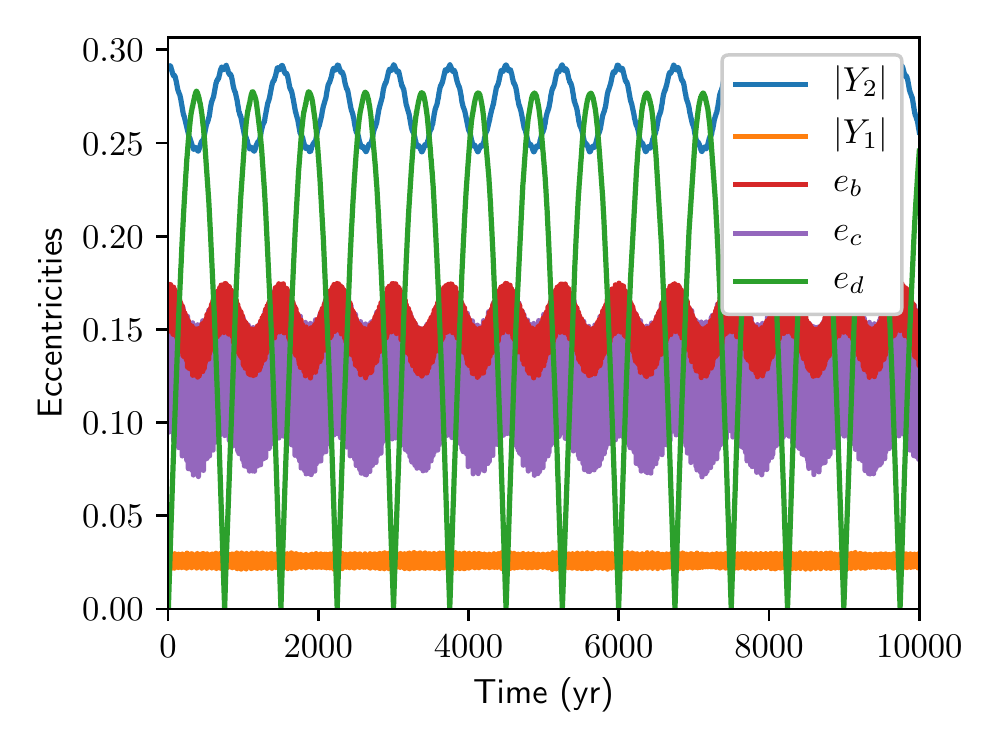}
	\caption{Eccentricities dynamical evolution over long time-scales when planet d is included. \label{fig:secevo}}
\end{figure}

We plot the eccentricities as well as $|Y_1|$ and $|Y_2|$ in Figure \ref{fig:secevo}.
We see that as in the case with only planets b and c, the eccentricities $e_b$ and $e_c$ evolve very rapidly with the roughly 6 yr period seen in section \ref{sec:shorttermecc} while $|Y_1|$ is almost constant.
We checked that the resonance is indeed preserved in this case during the whole integration.
However, $|Y_2|$ is no longer a constant and there are large AMD exchanges between planet d and the b-c pair.
Due to the smaller planet d mass and semimajor axis, its eccentricity rises to values around 0.37 and its mean value is 0.24.
We conclude that even starting with a circular orbit, planet d is largely coupled to the two outer planets and therefore cannot be considered in isolation.

\section{Tidal decay during long-term evolution}
\label{sec:tides}
From the last section we have seen that K2-19's system is stable over long time-scales. But until now we have only taken into account the purely $N$-body gravitational interactions.
However, due to their eccentric orbits, the three planets are subject to tidal effects.
Indeed, the change of orbital distance leads to friction inside the planet and thus energy dissipation.
Tidal effects conserve the total angular momentum\footnote{In this analysis, we neglect the planets spin as well as tides raised on the star, that influences the stellar spin. The total \emph{orbital} angular momentum is thus conserved} \citep{Goldreich1966}.  The energy loss results in a circularization of the orbits and a decay of the semimajor axes.

Dissipative effects are of particular importance due to the estimated age of the system. 
Indeed, based on K2-19's rotation period, the star is older than 1 Gyr. 
Besides, an age of a few Gyr is compatible with the rotation rate and effective temperature (David Trevor, private communication: the star is more slowly rotating so more likely older than similar temperature 1 Gyr old stars).

\subsection{Low-eccentricity tidal migration}

\cite{Pu2019} proposed a formation mechanism for ultrashort-period planets (with an orbital period of around 1 day), that they called the low-eccentricity migration scenario \citep[see also][]{Mardling2007,Laskar2012}.
They showed that, in a multiplanetary system with slightly eccentric outer planets, the inner planet migrates very efficiently inward up until the point where it is decoupled due to the precession induced by the star oblateness and general relativity or if the outer planets run out of AMD.

Since planet b and c interact secularly with planet d only through the variation of $Y_2$, we will assume that the interaction can be reduced to a two planet case: planet d and an outer planet. This simplification does not affect the results since the resonance is not affected by a variation of $Y_2$ (see figure \ref{fig:secevo}).
We use the two planet model presented by \cite{Pu2019} to compute the effect of tides on planet d's orbit. 
It should be noted that considering the two outer planet as a single one is a conservative assumption as their interactions could lead to a faster migration \citep{Pu2019}.

Following the weak friction theory of equilibrium tides \citep{Darwin1880,Alexander1973,Hut1981,Pu2019}, the evolution of the planets' semimajor axes in presence of tides is given by the equation
\begin{align}
\frac{\dot{a}_k}{a_k} =& -1.9\times10^{-9} k_{2,k} \left(\frac{\Delta t_{L,k}}{100\text{ s}}\right)\left(\frac{e_k}{0.02}\right)^2\left(\frac{m_S}{M_\odot}\right)^2\nonumber\\
&\times\left(\frac{m_k}{M_\oplus}\right)^{-1}\left(\frac{R_k}{R_\oplus}\right)^5\left(\frac{a_k}{0.02 \text{ au}}\right)^{-8} \text{ yr}^{-1},
\label{eq:tidaldecay}
\end{align}
where $k_{2,k}$ is the tidal Love number, $\Delta t_{L,k}$ is the tidal lag time and $R_k$ is the radius for planet $k$.
We note that the decay time-scale depends on the eccentricity squared. 
The decay is fast at moderate eccentricity and slows down as the orbit becomes more and more circular.
While the eccentricities of the outer planets are not extremely high, the validity of the limitation to leading order in eccentricities in eq. \eqref{eq:tidaldecay} should be discussed. 
From \citep{Hut1981}, we remark the the additional terms at order $e^4$ and $e^6$ accelerate the dissipation. Besides, while relevant at the beginning (when $e_b\simeq0.2$) the corrections become negligible for eccentricities closer to 0.1.
It results that the equation \eqref{eq:tidaldecay} gives a conservative decay rate and we do not include higher order corrections given the other uncertainties on the system such as on the tidal lag times.

The tides become less important for the farthest planets because of  the steep $a_k^{-8}$ dependency.
In principle tidal dissipation in the two outer planets should be considered. 
However, the tidal dissipation in large planets is not well constrained \citep{Ogilvie2014}. 
We thus follow the conservative assumption made in \cite{Pu2019} to neglect tides affecting planets b and c.

The decay rate also depends on the two coefficients $k_{2,d}$, of order unity, and $\Delta t_{L,d}$.
Using this parametrized formalism allows us to study the dissipation while remaining agnostic on the actual physical mechanisms at play. 
We show below that the results are compatible with a large range of values for the coefficients.
As \cite{Pu2019}, we take $k_{2,d}=1$. The tidal lag $\Delta t_{L,d}$ is inversely proportional to the planet's quality factor 
\begin{equation}Q_d = \frac{P_d}{4\upi\Delta t_{L,d}}.\end{equation}
Planet d is terrestrial so its tidal lag is close to 100 s \citep{Goldreich1966,Pu2019}, which corresponds to a quality factor close to 170. Using values close to the Solar system terrestrial planets is common in the field and motivated by studies of the viscoelastic response of planets to tidal deformations \citep{Correia2014,Efroimsky2014,Makarov2014}. 
In our analysis, we choose to draw planet d tidal lag time from a log-uniform distribution with boundaries  $50 \text{ s}\leq\Delta t_{L,d}\leq 500 \text{ s}$.

We first detail our model for the long-term evolution of the system and then discuss the possible outcomes.
As shown in section \ref{sec:secular}, the eccentricity of planet d is driven by its secular coupling with the outer planets.
For planet d, we replace in eq. \eqref{eq:tidaldecay} the eccentricity $e_d$ by a forced eccentricity $e_{d,\text{forced}}$ given by the secular coupling.
The forced eccentricity can be estimated as a function of planets b and c eccentricities. We use eq.~(40) from \cite{Pu2019},
\begin{equation}
e_{d,\text{forced}} = \frac{\nu_{d,b}}{\omega_{b,d}+\omega_{d,\text{gr}}+\omega_{d,\text{tide}}} e_b,
\label{eq:eforced}
\end{equation}
where $\nu_{d,b}$ and $\omega_{b,d}$ correspond to secular interaction terms between b and d, and $\omega_{d,\text{gr}}$ and $\omega_{d,\text{tide}}$ are respectively the apsidal precession of planet d due to general relativity and tides.
While planet d is close to its current position, $a_d\simeq 0.035$ au, the ratio $e_{d,\text{forced}}/e_b$ is close to 0.5.  For shorter orbits ($a_d\la 0.01$ au), the ratio sharply decays as general relativity and tidal precession become significant and decouple planet d from the outer planets, effectively stopping the tidal migration.
The secular planet interactions are computed using Lagrange-Laplace theory. 
We do not include higher order terms despite the moderate eccentricities.
Indeed, as we see in figure \ref{fig:secevo}, planet d eccentricity is well coupled to the pair b-c in numerical simulations. Higher order corrections will most likely give a more accurate coupling but eq. \eqref{eq:eforced} reproduces qualitatively the observed behaviour for moderate eccentricities and is accurate once the eccentricities become smaller.

\begin{figure}
	\includegraphics[width=\linewidth]{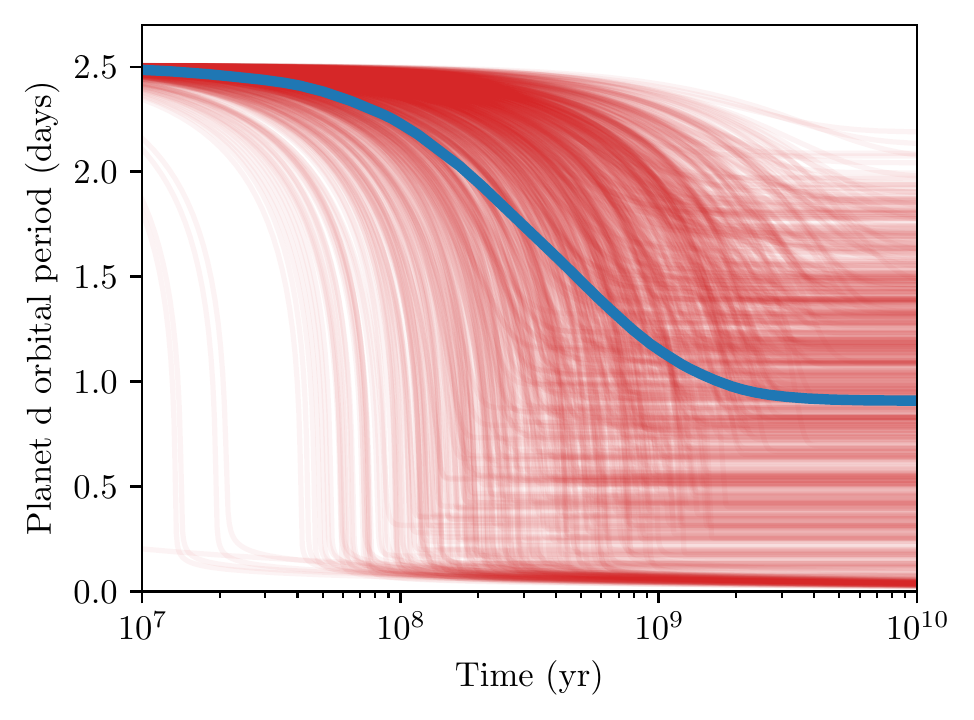}
	\caption{Planet d period evolution due to low-eccentricity tidal migration for 1,000 initial conditions drawn from the three planet best fit. The blue curve is the average value. 
		The spread is explained by the range of tidal lag times explored, the uncertainties on planet d's mass and the system's initial total AMD.\label{fig:period_decay}}
\end{figure} 

Finally, the outer planets eccentricities evolve as planet d migrates because tides raised on planets conserve the total orbital angular momentum\footnote{This is true as long as the planet spins are negligible with respect to the orbital angular momentum}.
One can estimate $e_b$ as a function of the new semimajor axis of planet d, and the initial values for $e_b$, $a_d$ and $a_b$
\begin{equation}
e_b^2 = 1- \left(1-\frac{C_0-(\Lambda_{d,0}-\Lambda_d)}{\Lambda_d+\Lambda_b+\Lambda_c}\right)^2,
\label{eq:ebAMD}
\end{equation}
where $C_0$ is the total initial AMD of the system and $\Lambda_{k,0}=m_k\sqrt{\Gr m_Sa_{k,0}}$ is the initial circular angular momentum of planet $k$. To obtain expression \eqref{eq:ebAMD}, we assumed that  planets b and c see no variation of their semimajor axis  due to tidal migration and that $e_c$ and $e_b$ were equal, which is reasonable in first approximation due to the resonance.
By inserting \eqref{eq:eforced} and \eqref{eq:ebAMD} into \eqref{eq:tidaldecay}, we obtain a differential equation for  $a_d$ that gives results comparable to a secular complete integration \citep{Pu2019}.

We draw 1,000 initial conditions from the posterior distribution of \cite{Petigura2020}. 
We plot in figure \ref{fig:period_decay} the evolution of the period of planet d over 10 Gyr.
The individual evolutions are in red while the thick blue line corresponds to the averaged value.

The final orbital periods extend from 0 d (where planet d would be consumed by the star) to almost 2 d.
It should be noted that for periods smaller than 0.4 d (semimajor axis of about 0.01 au), the decay is expected to be faster due to stellar tides that are neglected in this analysis. The typical outcome is the formation of an ultrashort-period planet with an orbit of about a day.
The final orbit of planet d is mainly determined by the mass of planet d and the initial AMD. For larger AMD or smaller mass $m_d$, the final orbit is shorter.

\begin{figure}
	\includegraphics[width=\linewidth]{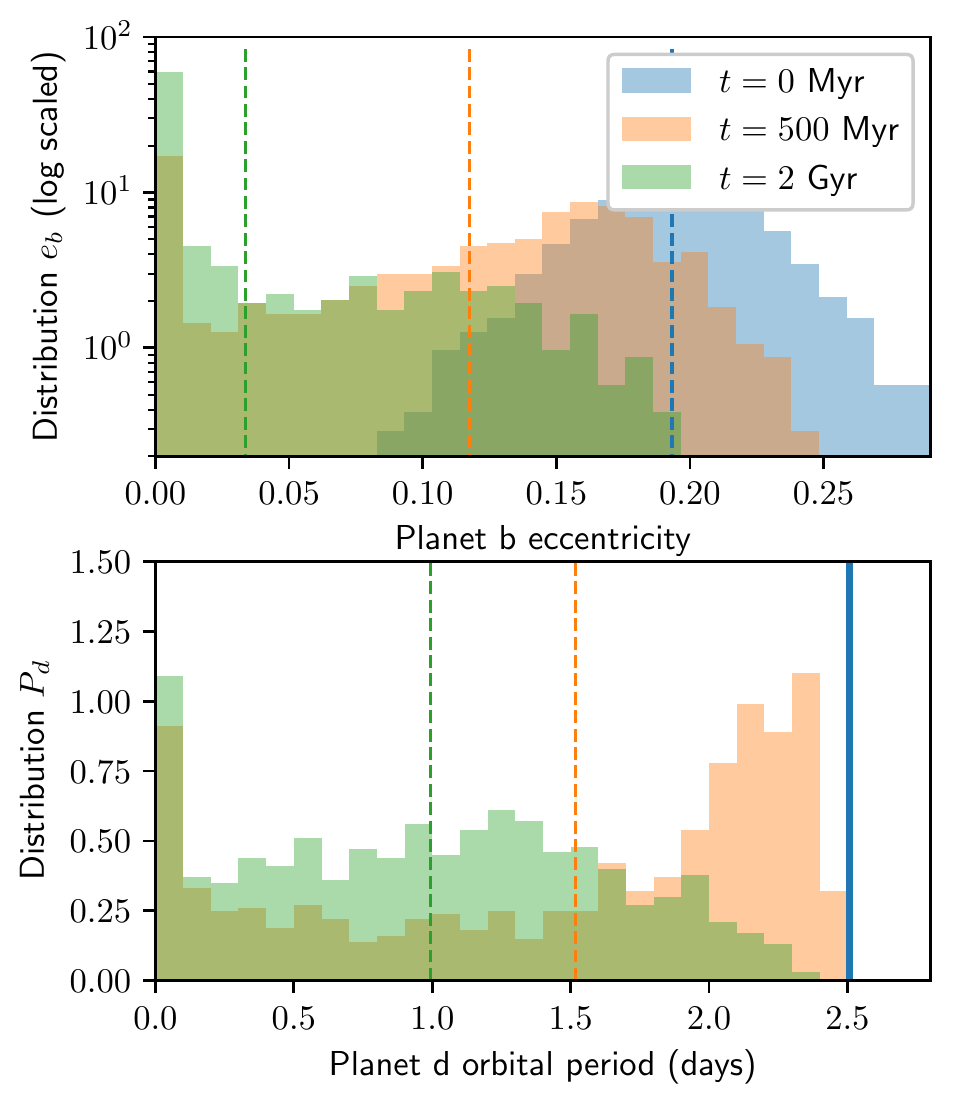}
	\caption{Distributionistribution of planet b eccentricity (top panel) and planet d orbital period (bottom panel) at various times. In the second panel, the initial period is represented by a blue thick line. The dashed vertical lines correspond to the mean values at the given time. Note the log scale in the first panel. \label{fig:pdfecc}}
\end{figure}

We plot in figure \ref{fig:pdfecc}, the resulting distribution of planet b eccentricity\footnote{We recall that planet c is assumed to have the same eccentricity as planet~b at all times.} and of planet d period at $t=0$ Myr, $t=500 $ Myr and at $t=2$ Gyr. 
We see that even after a few hundreds of Myr, the decay of the eccentricity and orbital period are general and significant.
After 2 Gyr, the eccentricity of planet b is smaller than 0.1 in 85\% of the simulations.
In the simulations where planet d does not migrate up to the star, the orbital decay is stopped because the AMD reservoir has been emptied, \emph{i.e.} the outer planets' orbits have been circularized.

More importantly, the final state is reached within a few hundred Myr. We define the half-decay time  $T_{\text{hf}}$ as the time such that planet d has undergone half of its orbital decay over 10 Gyr. The median half decay time is  ${T_{\text{hf}} = 472 \text{ Myr}}$ and 80\% of the initial conditions have  ${T_{\text{hf}} < 1 \text{ Gyr}.}$

\subsection{Tidal decay in the past history of the system}

We also considered the case where planet d started on a wider orbit and is currently experiencing tidal decay.
We run the same model but with planet d starting with a period of 3.5 d, correcting the total AMD such that the system keeps the same total angular momentum.
This initial period is the largest one before planet d's orbit crosses planet b's.
After 2 Gyr, only 30\% of the systems have planet d with a period larger than 2.4 d and $e_b>0.1$.
The systems compatible with today's observations give a constrain on planet d tidal lag.
The $1\sigma$ upper limit is $\Delta t_{L,d}^{86\%} = 130$ s (which corresponds to a quality factor of 180 at the current orbit).
The constraints become stronger if the system is older than 2 Gyr.

It results that it may be possible that the observed system is on its way to circularize in the next billion of years.
Such a scenario necessitates a very particular initial configuration where planet d is  originally on an orbit at the limit of instability.

\section{Truly eccentric? Tensions between the observations and the theory}
\label{sec:bias}

The observations for this system are very precise and from multiple sources. 
The planets were detected thanks to  the photometry from the K2 campaign \citep{Armstrong2015,Sinukoff2016}.
The orbital elements and masses are also constrained by RV data and TTV obtained 2 years after the \emph{K2} campaign.
\cite{Petigura2020} performs a photodynamical fit that forward model the lightcurve.
Given the posterior distribution obtained, a close to circular, three planet model is strongly ruled out.
Besides, the eccentricity cannot be attributed to a bias in the fitting model.
Indeed, the eccentricities and periapsis argument are parametrized as $\{\sqrt{e}\cos \omega,\sqrt{e}\sin\omega\}$ such that the prior is uniform in eccentricity \citep{Eastman2013}.
The eccentricity priors as well as the photodynamical modelling have been used in a number of previous studies.
In particular, if close to circular solutions provided a fit as good as the eccentric ones, the photodynamical model would have selected them.

In the limit of low eccentricities, the TTV signal is mainly affected by the distance to the nominal resonance ${\delta = \frac{2P_c}{3P_b}-1}$ and the variable $Y_1$ \citep{Lithwick2012a,Hadden2016}, but almost not by $Y_2$.
This is the reason why the constraint on $Y_1$ is much better than the one on $Y_2$.
Note that the formalism from \cite{Lithwick2012a} is developed for close but out of resonance planets. While $Y_1$ has a much greater effect on the TTVs than $Y_2$, the photodynamical model can extract more information from the TTV signal for moderate eccentricities. Indeed, we can show that the value of $Y_2$ as a strong influence on the obtained TTVs.
We draw 20 systems from the MCMC posterior and compute their transit times in numerical simulations.
We then refit the observed transits while forcing $Y_2$ to remain small ($\lesssim 0.03$) through numerical simulations and keep the best fit obtained.
In practice we perform a non linear least square fit on the transit times using a cost function defined as $\chi^2 = \frac{1}{2}\sum_j\left(T_{j}^{\mathrm{mod}}-T_{j}^{\mathrm{obs}}\right)^2/\sigma_j^2$. 
We force $Y_2$ to remain small by performing a first fit after adding a term in the cost function proportional to $|Y_2|^2$. 
We then take the result of this first fit as an initial condition to a second fit without penalization.

We checked that the close to circular systems were also inside the 3:2 MMR.
Note that this experiment's purpose is only to highlight the influence of $Y_2$ on the TTVs.
In particular, we do not take into account the information from the full photometry from \emph{K2}.
We take into account the RV constraints on the planet masses \citep[section 4.1.,][]{Petigura2020}\footnote{Fitting the transits without constraining the masses leads to systems where planet c mass is of order $20$ \Me whereas the RVs give an upper mass of 10.2 \Me (at 95\% confidence). The fit without the constraints were not leading to a significant cost improvement.}.

\begin{figure}
	\includegraphics[width=\linewidth]{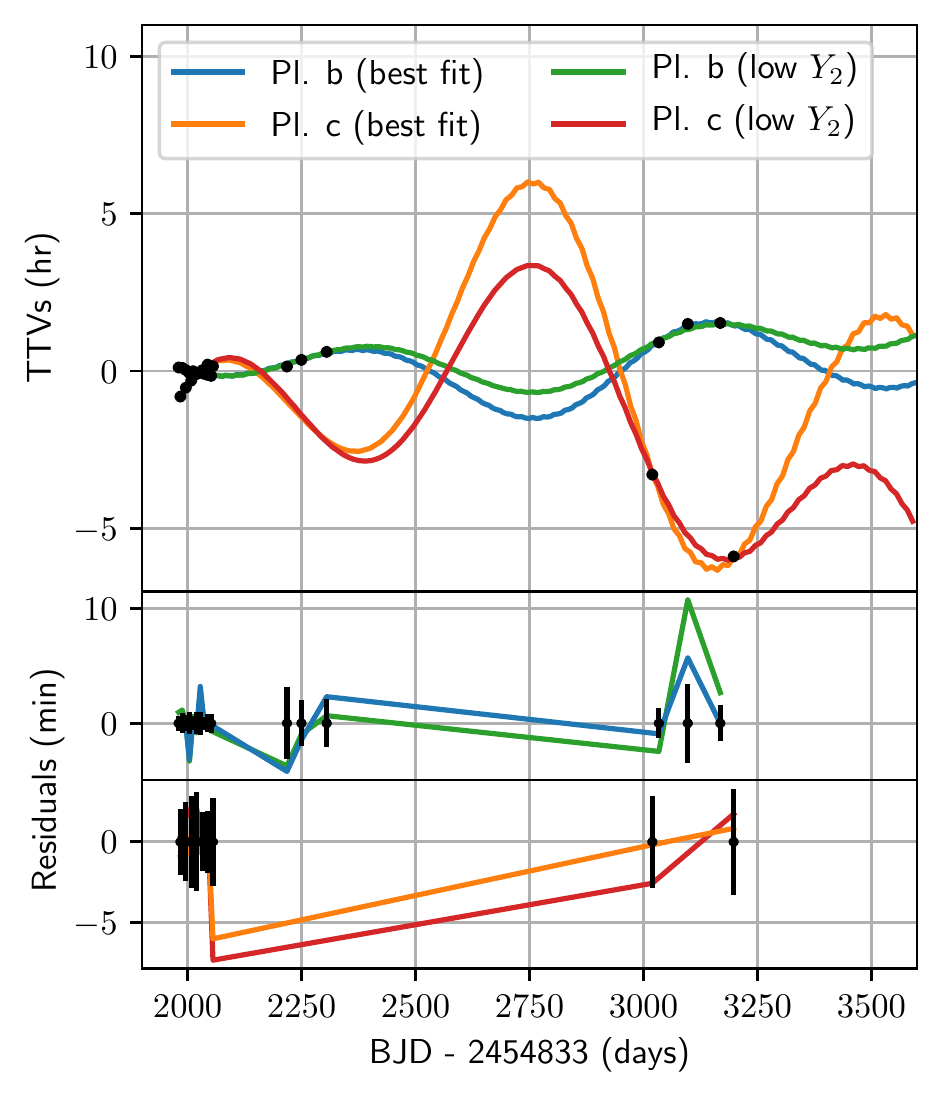}
	\caption{Synthetic TTV fits to the transit times. The "best fit" curves (blue for planet b, orange for planet c) correspond to the best fit where the eccentricities are not constrained. These systems have moderate eccentricities. The low $Y_2$ TTVs (green for planet b, red for planet c), result from a fit to the transit times while constraining the value of $Y_2$ (see text). The pair b-c remains inside the MMR but the orbits are very close to circular. The observed transit times are plotted as black dots. In the bottom panels, we give the residuals to the observed transit times (in min).\label{fig:TTV}}
\end{figure}

For the eccentric system and its close to circular counterpart, we plot in the first panel of Figure \ref{fig:TTV}, the synthetic transit times as well as the observed times from the \emph{K2} campaign\footnote{The \emph{K2} times and the associated errors were estimated in \cite{Narita2015}.} and from Table 1 of \citep{Petigura2020}.
On the two last panels, we plot the residuals to the observed times.
As expected, we see that the synthetic TTVs for the eccentric systems are consistent with the results obtained by \citep{Petigura2020}.
We note that the solutions where $Y_2$ was forced to remain small fit the transit times.
However, we see that the two sets of curves behave qualitatively differently.
We can also observe that the residuals are significantly larger for the close to circular fits with respect to the errors on the transit times.
In particular, the eccentric fit is compatible with the final set of \emph{Spitzer} data for planet b, whereas the transit times for the circular fit are at least 2 $\sigma$ away from the observations.
It shows that such follow-up measurements, taken long after the initial planet discoveries are critical to the characterization of the system.
Quantitatively, the goodness of fit of the eccentric, unconstrained fit is $\chi^2 = 17.7$ whereas it is $\chi^2 = 27.3$ for the close to circular ones.

This small experiment confirms the results from the photodynamical modelling: the current observations favour eccentric orbits over close to circular ones.
It should also be noted that the photodynamical modeling of the \emph{K2} lightcurve done in \citep{Petigura2020} also incorporates information on the transit durations  that is independent on the $Y_1$ measurement \citep{Kipping2010}.
While \citep{Petigura2020} ruled out $Y_2 = 0$ at 4$\sigma$ formal significance, we acknowledge the possibility of errors due to model-misspecification. The fact that $Y_2$ factors so critically into the system's dynamical interpretation motivates additional observations by the exoplanet community.

\cite{Petigura2020} already noticed that the architecture of the system was puzzling despite it being stable in numerical simulations.
However, it appears from the previous section that the tidal dissipation make the current system's architecture harder to explain.
Besides, the formation of the system remains unexplained.
So one needs to explain how to fit the observations while ensuring that the system configuration can be observed after a few Gyr.

\subsection{The formation challenge}

We showed in section \ref{sec:32res} that K2-19 b and c are trapped in the 3:2 MMR with a small libration amplitude.
Capture into MMR generally emerges from dissipative effects leading to convergent migration \citep{Batygin2015}.
The most common mechanism is migration within the protoplanetary disc \citep[\emph{e.g.}][]{Cresswell2008,Pichierri2019}.
Capture can also occurs due to convergent tidal migration \citep{Papaloizou2018}.
However, both disc and tidal migration show shorter time-scales for eccentricity damping  than for change in orbital period.
Systems are thus capture close to circular orbits, an increase of the eccentricity while the system is in the resonance typically leads to an anti-aligned configuration as pointed out by \cite{Petigura2020}.

In order to explain the present configuration, one has to imagine a mechanism where the planet's eccentricity vectors are not damped to zero but to a common value.
In this case, the capture can occur in the aligned configuration since the resonant dynamics and the capture mainly depend on $Y_1$ \citep[eq. 22 and fig. 7]{Batygin2015}.
Migration in eccentric disc has been studied theoretically \citep{Papaloizou2002}, but eccentric discs arise only in the presence of large planets \citep{Teyssandier2016} or for circumbinary discs. 
Moreover, eccentricities close to 0.2 leads to an outward migration \citep{DAngelo2006}, which seems at odds with the short period of planet b and c.
In any way, it is clear that the formation of the three planets around K2-19 remains a challenge that does not fit the classical scenarios.

\subsection{New constraints from tidal dissipation}

We have shown in section \ref{sec:tides} that taking into account tides is critical to understand the long-term evolution of this system.
Indeed, the best fit orbital solution is stable in the presence of purely gravitational interactions over long time-scales. 
However, the secular coupling between planet d and the resonant pair (see \ref{sec:secular}) leads to a strong tidal dissipation in the inner planet.
Over a few hundreds Myr, the outer planets' AMD is depleted and planet d experiences a period decay.
If the system had truly formed as we see it today, we would expect to observe the outer planets on circular orbits and the inner one on a shorter orbit.

Yet, the current system configuration might be explained in presence of tidal decay. 
However, it implies that planet d orbital period was originally larger ( up to 3.5 d) and that the three planets started on eccentric orbits with planet d and b at the limit of the orbit crossing.
It also requires that the tidal lag  time of planet d is smaller than 130s (or that the quality factor is larger than 180), a value that is not incompatible with our understanding of tides in rocky bodies but that still  gives a constraint on the dissipation rate.
Lower quality factors are strongly ruled out.
Such a fine tuning of the original configuration is necessary to explain the observed architecture if the system only hosts three planets.

In this work, we have not taken into account the dissipation in planets b and c because of the poor constraints on tidal dissipation in gaseous planets.
However, works on on tidal dissipation for systems in MMR \citep[\emph{e.g.}][]{Delisle2012a,Delisle2014a} have shown that the resonance is often broken before the planets circularization.
Such studies could give an additional constraint on the system.

\subsection{A fourth planet?}

When it is hard to reconcile the observations to the theory, it is common to explore the possibility that the planets motion can be perturbed by an unseen companion.
From Le Verrier's work on Neptune to the recent Planet 9 hypothesis \citep{Batygin2019}, this approach is historically tied to the progresses in our understanding of the Solar System because of the very precise constraints on the planets' motion.
However, the method have been applied with success also in the context of exoplanet dynamics.
One can cite the example of {Kepler-56}  where the 40 deg obliquity of the two transiting planets is due to a distant planetary companion \citep{Huber2013,Otor2016}.
But the most obvious application to exoplanets is the TTV method itself that allows us to find non-transiting planets, most of the time in resonance with a transiting one \citep[Kepler-88, the "King of TTVs" is an example of a system where a non-transiting planet perturbing the motion of Kepler-88b,][]{Nesvorny2013}.

The photodynamical fit strongly reject solutions with three planet on circular orbits.
Nevertheless, it remains possible that the model is misspecified, which would be the case if, for instance, the TTVs are affected by a fourth planet in the system.
In this case, the orbits of the observed planets b, c and d might be close to circular at the expense of the addition of another planet in the resonant chain.
Indeed, a few Earth masses planet trapped in an inner resonance with planet b can have significant TTVs contribution while being non transiting or even not being detected due to its small radius.
We also point out that the TTV coverage of \cite{Petigura2020} is sparse, especially compared to TTV datasets from the prime \emph{Kepler} mission. 
This dataset had timing measurements over three distinct epochs (the \emph{K2} campaign and two sets from \emph{Spitzer}). 
We expect sparse datasets to be more susceptible to model misspecification errors and encourage additional transit time measurements.

\begin{figure}
	\includegraphics[width=1\linewidth]{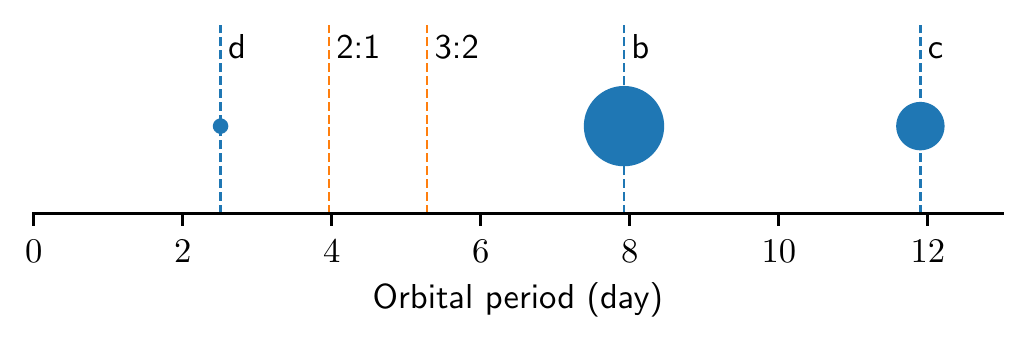}
	\caption{Architecture of the system K2-19 including the inner 2:1 and 3:2 MMR with planet b. The size of the planets is proportional to their radius.\label{fig:archi}}
\end{figure}

While it will necessitate more transit observations in the future to verify such a claim, it should be noted that it is possible to add a planet between  planet d and b without destabilizing the system.
Indeed, we show the positions of the planets alongside the resonances 2:1 and 3:2 with planet b  in figure \ref{fig:archi}.
Another planet in 3:2 or 2:1 resonance with planet b is consistent with the 'peas in a pod' pattern observed in the architecture of the \emph{Kepler} systems \citep{Weiss2018}.
Moreover, assuming this unseen planet is in a 2:1 resonance with planet b leads to a period ratio of 1.58 with planet d, which is just wide of the 3:2 resonance.
In other words, the whole system could have been placed into a four planet resonant chain during its formation before tidal effects broke the resonance with the innermost planet as it has been proposed by \cite{Millholland2019a} or \cite{Pichierri2019}.

\section{Conclusions and discussions}

In systems such as K2-19, the precise orbital parameters obtained by TTVs allow for rich dynamical studies that can provide a lot of insight onto the system history and configuration.
Such systems act as laboratories to test theories of formation and dynamical evolution.

Following their observations, \cite{Petigura2020} conclude that K2-19's two outer planets are very close to the 3:2 MMR but due to their large aligned eccentricity, the classical resonant angles are not librating.
A system with such a configuration is a challenge for classical planet formation scenarios.
Indeed, planets close to MMR tend to be captured due to type I migration during the disc lifetime \citep{Cresswell2008}. Besides, migration is very effective to damp planet's eccentricities.

We have shown in section \ref{sec:32res} that by considering the true resonant variables of the system that is a combination of the two complex eccentricities, the system rapid dynamics are well explained by the integrable first-order model.
Besides, the very small libration amplitude of the resonant variable (fig. \ref{fig:phaseportrait}) strengthens the classical scenario of a smooth capture through disc migration \citep{Batygin2015}.
We also see that the resonant variable is more constrained than the two eccentricities as it is expected for systems presenting TTVs \citep{Hadden2016}.
In particular, we want to highlight that in the context of orbital fit with such large eccentricities, the classical methods to spot a MMR such as monitoring the classical resonant angles (eq. \ref{eq:resanglesclassic}) is not reliable.
A resonant configuration can also exist even if the two orbits are not anti-aligned.

We then studied the entire system rather than only the two outer more massive planets.
The fitted configuration leads to a very strong secular coupling between the inner terrestrial planet d and the pair b and c.
The coupling does not disrupt the MMR but leads to eccentricities of the order of 0.35 for the inner planet.
In particular, this planetary system's architecture is long-lived in the presence of purely $N$-body interactions.

Because planet d orbits in about 2.5 d around its host star, it is subject to large tidal effects.
The tides raised on planet d tend to circularize its orbit at the expense of a period decay.
Following the low-eccentricity migration model presented in \citep{Pu2019}, we show that the two outer planets transfer their AMD to the inner planet through the secular coupling.
As a result the inner planet continues to decay up until the point where tidal and general relativity precession decouple it from the outer planets or if the outer planets run out of  AMD.
The typical time-scale of the process is less than 500 Myr whereas the system is expected to be billions of years old.
The system's current architecture remains compatible with the tidal decay if planet d started in a longer orbit, at the limit of orbit crossing with planet b.
Note that in the absence of planet d, it would be much harder to rule out the aligned eccentric resonant configuration due to the poor constraint we have on the dissipation onto gaseous planets.

Even though the photodynamical fit gives a configuration that can only be maintained for a short amount of time with respect to the system lifetime, the observations have to be explained.
While biases in eccentricity determination are known and have been quantified in RV observations \citep{Anglada-Escude2010,Hara2019}, no such study has been carried for TTVs systems.
One could speculate on the fact that another undiscovered planet in the system can affect the outer planets periods.
Indeed, in the picture of the "peas in a pod" systems \citep{Weiss2018}, the system is compatible with the presence of another planet between planet d and b.
If such a non-transiting planet could also be in resonance with planet b and c and lead to TTV unaccounted for.
However, the presence of an eventual fourth planet would need to be confirmed by more measurements of planet b and c transits.

The detailed analysis of the K2-19 system has revealed even richer dynamics than originally reported.
Taking into account the dissipative effects also appears to be crucial for the understanding of the system history.
Nevertheless, the system formation remains mostly unexplained. Future photometric or RV monitoring will be crucial to unveil the nature of this system.

\section*{Data availability}
The data and code used to generate the plots in this article will be shared on reasonable request to the corresponding author.

\section*{Acknowledgements}
The authors wish to thank the anonymous referee, Konstantin Batygin, Trevor David, Alexander Mustill and Gabriele Pichierri for helpful comments and discussions.
 M.D. and A.P. are supported by the project grant 2014.0017 ‘IMPACT’ from the Knut and Alice Wallenberg Foundation. A.J. and A.P. was supported by the European Research Council under ERC Consolidator Grant agreement 724687-PLANETESYS, the Swedish Research Council (grant 2018-04867), and the Knut and Alice Wallenberg Foundation (grants 2014.0017 and 2017.0287).

\bibliography{K2-19}{}
\bibliographystyle{aasjournal}

\appendix

\section{Mean Motion Resonance dynamics}
\label{app:ham}

In this section, we briefly derive the first-order integrable model for the $p$+1:$p$ MMR. The reader interested in a recent detailed analysis should refer to \cite{Batygin2013,Deck2013,Petit2017,Hadden2019}.
The model was initially developed by \cite{Henrard1983,Sessin1984,Henrard1986}.
We consider two planets of mass $m_1$ and $m_2$ orbiting a star of mass $m_0$ in a plane (the spatial case is treated similarly).
The Hamiltonian in democratic heliocentric coordinates is \citep[\emph{e.g.}][]{Morbidelli2002}
\begin{equation}
\H = \sum_{k=1}^{2}\frac{||\tr_k||^2}{2m_k} - \frac{\mu m_k}{r_k} + \frac{||\tr_1+\tr_2||^2}{2m_0} -\frac{\Gr m_1m_2}{r_{12}},
\label{eq:ham2p}
\end{equation}
where $\tr_k = m_k \dot{\r}_k$ is the barycentric momentum, $\r_k$ the heliocentric position $\Gr$ is the gravitational constant and $\mu =\Gr m_0$ .
We may express $\H$ in terms of the complex Poincaré coordinates \citep[\emph{e.g.}][]{Laskar1990} 
\begin{align}
\Lambda_k &= m_k\sqrt{\mu a_k}, & \lambda_k;\nonumber\\
C_k &= \Lambda_k\left(1-\sqrt{1-e_k^2}\right) ,& -\varpi_k;\\
& \sqrt{C_k}e^{-\i\varpi_k} , &\i  \sqrt{C_k}e^{\i\varpi_k} ;\nonumber
 \end{align}
where $a_k$ is the semimajor axis, $e_k$ the eccentricity, $\lambda_k$ the mean longitude and $\varpi_k$ the longitude of periapsis of planet $k$. Note that the last two lines are redundant and both set of variables can be used to obtain the Hamiltonian equations.
Expressed in these variables, the Hamiltonian takes the form $\H_0+\epsilon\H_1$ where $\H_0$ is the Keplerian part \begin{equation}\H_0 = -\frac{\mu^2m_1^3}{2\Lambda_1^2}-\frac{\mu^2m_2^3}{2\Lambda_2^2}\end{equation} and $\epsilon\H_1$ is the perturbation part and depends on all the coordinates. The small parameter $\epsilon=(m_1+m_2)/m_0$ is introduced explicitly to emphasize the scale difference.

In order to obtain an integrable model for the $p+1$:$p$ MMR, we carry out a canonical transformation such that the new coordinates are \citep[\emph{e.g.}][]{Petit2017}
\begin{align}
\Gamma &= \frac{p+1}{p}\Lambda_1+\Lambda_2, & \theta_\Gamma & = p(\lambda_1-\lambda_2),\nonumber;\\
G &= \Lambda_1+\Lambda_2-C_1-C_2, & \theres& = (p+1)\lambda_2-p\lambda_1;\label{eq:lintrans}\\
x_k &= \sqrt{C_k}e^{\i(\varpi_k-\theres)} ,& \i\bar{x}_k&= \i \sqrt{C_k}e^{-\i(\varpi_k-\theres)}; \nonumber
\end{align}
$\Gamma$ has been called in the literature the scaling factor \cite{Sessin1984}, $G$ is the total angular momentum.
Because of the d'Alembert relations, the Hamiltonian does not depend explicitly on $\theres$ (\emph{i.e.} the angular momentum is conserved).
The two resonant angles are  $\varpi_k-\theres$ (see eq. \ref{eq:resanglesclassic}). We then expand the perturbation $\epsilon\H_1$ in power series of $x_k$ and Fourier series of $\theta_\Gamma$  and only keep the first-order terms.
The next step consists in averaging the motion over the fast angle $\theta_\Gamma$.
This operation is also a canonical transformations and the new coordinates are $\epsilon$ close to the old ones. They correspond to the average coordinates over a  Keplerian orbit.
The resulting Hamiltonian no longer depends on $\theta_\Gamma$. 
As a result, $\Gamma$ is a constant of the averaged system.
Note that the inverse transformation of eq. \eqref{eq:lintrans} allow to express $\Lambda_k$ as a function of the total AMD, $C$ and the constants of motion $\Gamma$ and $G$.

The perturbation $\epsilon\H_1$ then takes the form
\begin{equation}
\epsilon\H_1= \epsilon R_1( x_1+\bar{ x}_1)+ \epsilon R_2(x_2+\bar{x}_2),
\end{equation}
where
\begin{equation}
R_1= -\frac{\gamma}{1+\gamma}\frac{\mu^2m_2^3}{\Lambda_2^2}\frac{1}{2}\sqrt{\frac{2}{\Lambda_1}}\R_1(\alpha)\end{equation}
{and}
\begin{equation}
R_2= -\frac{\gamma}{1+\gamma}\frac{\mu^2m_2^3}{\Lambda_2^2}\frac{1}{2}\sqrt{\frac{2}{\Lambda_2}}\R_2(\alpha)
\end{equation}
with $\gamma=m_1/m_2$ and $\alpha =a_1/a_2$, 
\begin{align}
\R_1(\alpha)&=-\frac{\alpha}{4}\left(3\lc{3/2}{p}{\alpha}-2\alpha\lc{3/2}{p+1}{\alpha}-\lc{3/2}{p+2}{\alpha}\right),
\label{eq.coef_r1} \\
\R_2(\alpha)&=\frac{\alpha}{4}\left(3\lc{3/2}{p-1}{\alpha}-2\alpha\lc{3/2}{p}{\alpha}-\lc{3/2}{p+1}{\alpha}\right) +\frac{1}{2}\lc{1/2}{p}{\alpha}.
\label{eq.coef_r2}
\end{align}
The opposite signs of $\R_1$ and $\R_2$ should be noted.
In the two previous expressions, $\lc{s}{k}{\alpha}$ are the Laplace coefficients that can be expressed as
\begin{equation}
\lc{s}{k}{\alpha}=\frac{1}{\upi}\int_{-\pi}^{\pi}\frac{\cos(k\phi)}{\left(1-2\alpha\cos\phi+\alpha^2\right)^s}\mathrm{d}\phi
\label{eq.Lapcoef}
\end{equation}
for $k>0$. For $k=0$, a $1/2$ factor has to be added in the second-hand member of (\ref{eq.Lapcoef}). Note that for $p=1$ \emph{i.e.} the 2:1 MMR, a contribution from the indirect part (due to the reflex motion of the star) should be added \citep{Delisle2012a}.
Because we restrict the expansion to the first-order, all the coefficients are constant and evaluated at the Keplerian resonance, $\alpha=\alpha_0=(p/(p+1))^{2/3}$.

The integrable Hamiltonian is then obtained by a final transformation proposed by \cite{Sessin1984} and \cite{Henrard1986} and generalized by \cite{Hadden2019} to resonances of arbitrary order. Geometrically it consists in a rotation of the coordinates $x_1$ and $x_2$.
We define 
\begin{align}
y_1 &= \frac{R_1}{\sqrt{R_1^2+R_2^2}}x_1 +\frac{R_2}{\sqrt{R_1^2+R_2^2}}x_2\nonumber\\
y_2 &= \frac{R_2}{\sqrt{R_1^2+R_2^2}}x_1 -\frac{R_1}{\sqrt{R_1^2+R_2^2}}x_2.
\end{align}
Let us define $I_k = y_k\bar{y}_k$. $I_2$ now becomes a constant of motion and $C=I_1+I_2$.
$I_1$ and $I_2$ do not depend on the resonant angles but only on $\Delta\varpi$. Indeed one has
\begin{align}
I_1 &= \frac{1}{R_1^2+R_2^2}\left(R_1^2C_1 + R_2^2 C_2 +2R_1R_2\sqrt{C_1C_2}\cos (\Delta \varpi)\right),\nonumber\\
I_2 &= \frac{1}{R_1^2+R_2^2}\left(R_2^2C_1 + R_1^2 C_2 -2R_1R_2\sqrt{C_1C_2}\cos (\Delta \varpi)\right).
\end{align}

Finally, we expand the Keplerian part close to the circular Keplerian resonance given the scaling factor $\Gamma$ and the angular momentum G
\begin{equation}
\H_0 = \frac{\mathcal{K}_2}{2}(C-\Delta G)^2,
\end{equation}
where 
\begin{equation}
\frac{\mathcal{K}_2}{2} = -\frac{3n_2}{2\Lambda_2}(p+1)^2\frac{\gamma+\alpha}{\alpha},
\end{equation}
$n_2=2\pi/P_2$ being the mean motion and 
\begin{equation}
\Delta G = \frac{\Gamma}{p+1}\frac{p\gamma+(p+1)\alpha_0}{\gamma+\alpha_0}-G.
\end{equation}
The integrable Hamiltonian plotted in figure \ref{fig:phaseportrait} has the form
\begin{equation}
	\H = \frac{\mathcal{K}_2}{2}(I_1+I_2-\Delta G)^2 + 2\epsilon\sqrt{R_1^2+R_2^2}\sqrt{I_1}\cos(\theta_1)
	\label{eq:ham}
\end{equation}
where $\theta_1$ is the argument of $y_1$. Such a Hamiltonian is called the second fundamental model of resonance.
We refer to the previously cited papers and to \citep{Ferraz-Mello2007} for a complete description of the dynamics.

When plotting the Hamiltonian level curves in Figure \ref{fig:phaseportrait}, we choose the value of the parameters from the short-term integration shown in grey.
This choice is motivated by the fact that the posterior values of $I_1$ and $I_2-\Delta G$ are correlated.
Indeed, while the uncertainties on $I_1$  are much smaller than the uncertainties on $I_2$, it turns out, that the uncertainties on the quantity $I_1+I_2-\Delta G$ are even smaller.
The ratio of the standard deviations of $I_1+I_2-\Delta G$ and $I_1$ is 0.07.
Since the shape of the resonance is affected by small variations of $I_2-\Delta G$, the values associated to the integration are more faithful to the actual dynamics.
We also emphasize that it is critical to compute $I_2$ with the exact expression and not the linear approximations for $\tilde{e}_k$ as the approximation leads to a significant deformation of the phase space.

As claimed in section \ref{sec:32res}, $\theta_2 = \arg(Y_2)$ precesses at the same frequency as $\theres$. 
Indeed, by definition of the canonical variables, we have
\begin{equation}
\dot{\theta}_2 = \frac{\partial\H}{\partial I_2} = \mathcal{K}_2(C-\Delta G) = \frac{\partial\H}{\partial G} = \dot{\theta}_\mathrm{res}.
\end{equation}
The frequency can be related to the distance to the Keplerian resonance $\delta = pn_1/((p+1)n_2)-1$ and we have $\dot{\theta}_2 = -n_2\delta(p+1)$.

\end{document}

%% file: val_sys.tex
\newcommand{\sys}[1]{%
\IfEqCase{#1}{%
{steff}{5322}%
{steff_err}{100}%
{slogg}{4.51}%
{slogg_err}{0.08}%
{smet}{0.06}%
{smet_err}{0.05}%
{smass}{0.88}%
{smass_err}{0.03}%
{srad}{0.82}%
{srad_err}{0.04}%
{phot-gp-eta1}{0.4}%
{phot-gp-eta1_err}{0.1}%
{phot-gp-eta2}{34}%
{phot-gp-eta2_err}{5}%
{phot-gp-eta3}{20.4}%
{phot-gp-eta3_err}{0.3}%
{phot-gp-eta4}{0.49}%
{phot-gp-eta4_err}{0.06}%
{rv-per1}{7.92075}%
{rv-per2}{11.89804}%
{rv-per3}{2.50856}%
{rv-tc1}{1980.379}%
{rv-tc2}{1984.305}%
{rv-tc3}{1975.921}%
{rv-k1}{11.4}%
{rv-k1_err1}{1.7}%
{rv-k1_err2}{-1.7}%
{rv-k1_fmt}{$11.4 \pm 1.7$}%
{rv-k2}{0.4}%
{rv-k2_err1}{1.5}%
{rv-k2_err2}{-1.5}%
{rv-k2_fmt}{$0.4 \pm 1.5$}%
{rv-k3}{-0.2}%
{rv-k3_err1}{1.0}%
{rv-k3_err2}{-1.0}%
{rv-k3_fmt}{$-0.2 \pm 1.0$}%
{rv-mpsini1}{33}%
{rv-mpsini1_err1}{5}%
{rv-mpsini1_err2}{-5}%
{rv-mpsini1_fmt}{$33 \pm 5$}%
{rv-mpsini2}{1}%
{rv-mpsini2_err1}{5}%
{rv-mpsini2_err2}{-5}%
{rv-mpsini2_fmt}{$1 \pm 5$}%
{rv-mpsini3}{-0}%
{rv-mpsini3_err1}{2}%
{rv-mpsini3_err2}{-2}%
{rv-mpsini3_fmt}{$-0 \pm 2$}%
{rv-musini1}{111}%
{rv-musini1_err1}{17}%
{rv-musini1_err2}{-17}%
{rv-musini1_fmt}{$111 \pm 17$}%
{rv-musini2}{5}%
{rv-musini2_err1}{17}%
{rv-musini2_err2}{-17}%
{rv-musini2_fmt}{$5 \pm 17$}%
{rv-musini3}{-1.1}%
{rv-musini3_err1}{6.7}%
{rv-musini3_err2}{-6.7}%
{rv-musini3_fmt}{$-1.1 \pm 6.7$}%
{rv-gamma_j}{-2}%
{rv-gamma_j_err1}{2}%
{rv-gamma_j_err2}{-2}%
{rv-gamma_j_fmt}{$-2 \pm 2$}%
{rv-dvdt}{5.9}%
{rv-dvdt_err1}{2.4}%
{rv-dvdt_err2}{-2.4}%
{rv-dvdt_fmt}{$5.9 \pm 2.4$}%
{rv-gp_amp}{7.4}%
{rv-gp_amp_err1}{2.2}%
{rv-gp_amp_err2}{-2.2}%
{rv-gp_amp_fmt}{$7.4 \pm 2.2$}%
{rv-gp_explength}{35.0}%
{rv-gp_explength_err1}{4.7}%
{rv-gp_explength_err2}{-4.7}%
{rv-gp_explength_fmt}{$35.0 \pm 4.7$}%
{rv-gp_per}{20.3}%
{rv-gp_per_err1}{0.3}%
{rv-gp_per_err2}{-0.3}%
{rv-gp_per_fmt}{$20.3 \pm 0.3$}%
{rv-gp_perlength}{0.46}%
{rv-gp_perlength_err1}{0.07}%
{rv-gp_perlength_err2}{-0.07}%
{rv-gp_perlength_fmt}{$0.46 \pm 0.07$}%
{rv-mpsini2_p95}{10.2}%
{rv-mpsini3_p95}{3.5}%
{pd-mstar}{0.88}%
{pd-mstar_err1}{0.03}%
{pd-mstar_err2}{-0.03}%
{pd-mstar_fmt}{$0.88 \pm 0.03$}%
{pd-rstar}{0.82}%
{pd-rstar_err1}{0.03}%
{pd-rstar_err2}{-0.03}%
{pd-rstar_fmt}{$0.82 \pm 0.03$}%
{pd-masse1}{5.1}%
{pd-masse1_err1}{2.4}%
{pd-masse1_err2}{-2.4}%
{pd-masse1_fmt}{$5.1 \pm 2.4$}%
{pd-masse2}{32}%
{pd-masse2_err1}{2}%
{pd-masse2_err2}{-2}%
{pd-masse2_fmt}{$32 \pm 2$}%
{pd-masse3}{10.8}%
{pd-masse3_err1}{0.6}%
{pd-masse3_err2}{-0.6}%
{pd-masse3_fmt}{$10.8 \pm 0.6$}%
{pd-per1}{2.5081}%
{pd-per1_err1}{0.0002}%
{pd-per1_err2}{-0.0002}%
{pd-per1_fmt}{$2.5081 \pm 0.0002$}%
{pd-per2}{7.9222}%
{pd-per2_err1}{0.0001}%
{pd-per2_err2}{-0.0001}%
{pd-per2_fmt}{$7.9222 \pm 0.0001$}%
{pd-per3}{11.8993}%
{pd-per3_err1}{0.0008}%
{pd-per3_err2}{-0.0008}%
{pd-per3_fmt}{$11.8993 \pm 0.0008$}%
{pd-tc1}{2021.0726}%
{pd-tc1_err1}{0.0018}%
{pd-tc1_err2}{-0.0018}%
{pd-tc1_fmt}{$2021.0726 \pm 0.0018$}%
{pd-tc2}{2027.9023}%
{pd-tc2_err1}{0.0002}%
{pd-tc2_err2}{-0.0002}%
{pd-tc2_fmt}{$2027.9023 \pm 0.0002$}%
{pd-tc3}{2020.0007}%
{pd-tc3_err1}{0.0004}%
{pd-tc3_err2}{-0.0004}%
{pd-tc3_fmt}{$2020.0007 \pm 0.0004$}%
{pd-secosw1}{0.00}%
{pd-secosw1_err1}{0.00}%
{pd-secosw1_err2}{-0.00}%
{pd-secosw1_fmt}{$0.00 \pm 0.00$}%
{pd-secosw2}{0.02}%
{pd-secosw2_err1}{0.06}%
{pd-secosw2_err2}{-0.06}%
{pd-secosw2_fmt}{$0.02 \pm 0.06$}%
{pd-secosw3}{0.04}%
{pd-secosw3_err1}{0.04}%
{pd-secosw3_err2}{-0.04}%
{pd-secosw3_fmt}{$0.04 \pm 0.04$}%
{pd-sesinw1}{0.00}%
{pd-sesinw1_err1}{0.00}%
{pd-sesinw1_err2}{-0.00}%
{pd-sesinw1_fmt}{$0.00 \pm 0.00$}%
{pd-sesinw2}{-0.44}%
{pd-sesinw2_err1}{0.04}%
{pd-sesinw2_err2}{-0.04}%
{pd-sesinw2_fmt}{$-0.44 \pm 0.04$}%
{pd-sesinw3}{-0.46}%
{pd-sesinw3_err1}{0.03}%
{pd-sesinw3_err2}{-0.03}%
{pd-sesinw3_fmt}{$-0.46 \pm 0.03$}%
{pd-inc1}{90.8}%
{pd-inc1_err1}{0.7}%
{pd-inc1_err2}{-0.7}%
{pd-inc1_fmt}{$90.8 \pm 0.7$}%
{pd-inc2}{91.5}%
{pd-inc2_err1}{0.1}%
{pd-inc2_err2}{-0.1}%
{pd-inc2_fmt}{$91.5 \pm 0.1$}%
{pd-inc3}{91.1}%
{pd-inc3_err1}{0.1}%
{pd-inc3_err2}{-0.1}%
{pd-inc3_fmt}{$91.1 \pm 0.1$}%
{pd-Omega1}{0.0}%
{pd-Omega1_err1}{0.0}%
{pd-Omega1_err2}{-0.0}%
{pd-Omega1_fmt}{$0.0 \pm 0.0$}%
{pd-Omega2}{0.0}%
{pd-Omega2_err1}{0.0}%
{pd-Omega2_err2}{-0.0}%
{pd-Omega2_fmt}{$0.0 \pm 0.0$}%
{pd-Omega3}{-7.4}%
{pd-Omega3_err1}{0.8}%
{pd-Omega3_err2}{-0.8}%
{pd-Omega3_fmt}{$-7.4 \pm 0.8$}%
{pd-omegadeg1}{0}%
{pd-omegadeg1_err1}{0}%
{pd-omegadeg1_err2}{-0}%
{pd-omegadeg1_fmt}{$0 \pm 0$}%
{pd-omegadeg2}{-88}%
{pd-omegadeg2_err1}{8}%
{pd-omegadeg2_err2}{-8}%
{pd-omegadeg2_fmt}{$-88 \pm 8$}%
{pd-omegadeg3}{-85}%
{pd-omegadeg3_err1}{6}%
{pd-omegadeg3_err2}{-6}%
{pd-omegadeg3_fmt}{$-85 \pm 6$}%
{pd-omegadiffdeg}{2}%
{pd-omegadiffdeg_err1}{2}%
{pd-omegadiffdeg_err2}{-2}%
{pd-omegadiffdeg_fmt}{$2 \pm 2$}%
{pd-e1}{0.00}%
{pd-e1_err1}{0.00}%
{pd-e1_err2}{-0.00}%
{pd-e1_fmt}{$0.00 \pm 0.00$}%
{pd-e2}{0.20}%
{pd-e2_err1}{0.03}%
{pd-e2_err2}{-0.03}%
{pd-e2_fmt}{$0.20 \pm 0.03$}%
{pd-e3}{0.21}%
{pd-e3_err1}{0.03}%
{pd-e3_err2}{-0.03}%
{pd-e3_fmt}{$0.21 \pm 0.03$}%
{pd-ror1}{0.0124}%
{pd-ror1_err1}{0.0004}%
{pd-ror1_err2}{-0.0004}%
{pd-ror1_fmt}{$0.0124 \pm 0.0004$}%
{pd-ror2}{0.0777}%
{pd-ror2_err1}{0.0006}%
{pd-ror2_err2}{-0.0006}%
{pd-ror2_fmt}{$0.0777 \pm 0.0006$}%
{pd-ror3}{0.0458}%
{pd-ror3_err1}{0.0004}%
{pd-ror3_err2}{-0.0004}%
{pd-ror3_fmt}{$0.0458 \pm 0.0004$}%
{pd-c1}{0.4}%
{pd-c1_err1}{0.1}%
{pd-c1_err2}{-0.1}%
{pd-c1_fmt}{$0.4 \pm 0.1$}%
{pd-c2}{0.3}%
{pd-c2_err1}{0.2}%
{pd-c2_err2}{-0.2}%
{pd-c2_fmt}{$0.3 \pm 0.2$}%
{pd-prad1}{1.11}%
{pd-prad1_err1}{0.05}%
{pd-prad1_err2}{-0.05}%
{pd-prad1_fmt}{$1.11 \pm 0.05$}%
{pd-prad2}{7.0}%
{pd-prad2_err1}{0.2}%
{pd-prad2_err2}{-0.2}%
{pd-prad2_fmt}{$7.0 \pm 0.2$}%
{pd-prad3}{4.1}%
{pd-prad3_err1}{0.2}%
{pd-prad3_err2}{-0.2}%
{pd-prad3_fmt}{$4.1 \pm 0.2$}%
{pd-masseprec1}{5.07}%
{pd-masseprec1_err1}{2.40}%
{pd-masseprec1_err2}{-2.40}%
{pd-masseprec1_fmt}{$5.07 \pm 2.40$}%
{pd-masseprec2}{32.4}%
{pd-masseprec2_err1}{1.7}%
{pd-masseprec2_err2}{-1.7}%
{pd-masseprec2_fmt}{$32.4 \pm 1.7$}%
{pd-masseprec3}{10.80}%
{pd-masseprec3_err1}{0.57}%
{pd-masseprec3_err2}{-0.57}%
{pd-masseprec3_fmt}{$10.80 \pm 0.57$}%
{lopez-fenv2}{44}%
{lopez-fenv2_err1}{3}%
{lopez-fenv2_err2}{-3}%
{lopez-fenv2_fmt}{$44 \pm 3$}%
{lopez-mcore2}{18}%
{lopez-mcore2_err1}{1}%
{lopez-mcore2_err2}{-1}%
{lopez-mcore2_fmt}{$18 \pm 1$}%
{lopez-fenv3}{14}%
{lopez-fenv3_err1}{1}%
{lopez-fenv3_err2}{-1}%
{lopez-fenv3_fmt}{$14 \pm 1$}%
{lopez-mcore3}{9.4}%
{lopez-mcore3_err1}{0.5}%
{lopez-mcore3_err2}{-0.5}%
{lopez-mcore3_fmt}{$9.4 \pm 0.5$}%
{lopez-teq1}{1247}%
{lopez-teq1_err1}{38}%
{lopez-teq1_err2}{-38}%
{lopez-teq1_fmt}{$1247 \pm 38$}%
{lopez-teq2}{850}%
{lopez-teq2_err1}{26}%
{lopez-teq2_err2}{-26}%
{lopez-teq2_fmt}{$850 \pm 26$}%
{lopez-teq3}{742}%
{lopez-teq3_err1}{23}%
{lopez-teq3_err2}{-23}%
{lopez-teq3_fmt}{$742 \pm 23$}%
}[XX]%
}